\documentclass[a4paper,11pt]{article}
\usepackage{arxiv}
\usepackage[utf8]{inputenc}
\usepackage[T1]{fontenc}
\usepackage{hyperref}
\usepackage{url}
\usepackage{booktabs}
\usepackage{amsfonts}
\usepackage{graphicx}
\usepackage{doi}
\usepackage{xspace}
\usepackage{amsthm}
\usepackage{amsmath}
\usepackage[algo2e,noend,linesnumbered,ruled,vlined]{algorithm2e}
\usepackage{empheq}
\usepackage{subcaption}
\usepackage[table]{xcolor}
\usepackage{multirow}

\newcommand{\Z}{\mathbb{Z}}

\newcommand{\GBP}{\texttt{GBP}\xspace}
\newcommand{\GBPIP}{\texttt{GBP-IP}\xspace}

\newcommand{\ILP}{\texttt{ILP}\xspace}
\newcommand{\CSP}[1]{\texttt{CSP{#1}}\xspace}
\newcommand{\CMCP}{\texttt{CMCP}\xspace}
\newcommand{\CMCPIP}{\texttt{CMCP-IP}\xspace}
\newcommand{\BFF}{\texttt{BFF}\xspace}
\newcommand{\BFFd}{\texttt{BFF-d}\xspace}
\newcommand{\GDCA}{\texttt{GDCA}\xspace}
\newcommand{\PRYM}{\texttt{PRYM}\xspace}
\newcommand{\bigO}{\mathcal{O}}
\newcommand{\NPclass}{{\sf NP}}
\newcommand{\ie}{\textit{i.e.}}
\newcommand{\eg}{\textit{e.g.}}
\newcommand{\IP}{\texttt{IP}\xspace}

\newtheorem{theorem}{Theorem}
\newtheorem{problem}{Problem}
\newtheorem{conjecture}{Conjecture}
\newtheorem{proposition}{Proposition}

\title{Solving the Graph Burning Problem for Large Graphs\thanks{This work was supported in part by grants from: {\it Santander Bank}, Brazil; {\it Brazilian National Council for Scientific and Technological Development} (CNPq), Brazil, \#314293/2023-0, \#313329/2020-6; {\it São Paulo Research Foundation} (FAPESP), Brazil, \#2023/04318-7; \#2023/14427-8.}}

\date{April 25, 2024}

\author{
    \href{https://orcid.org/0000-0002-8967-8576}{\includegraphics[scale=0.06]{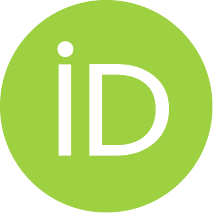}\hspace{1mm}Felipe de C. Pereira}\thanks{Corresponding author.} \\
    Institute of Computing\\
    University of Campinas\\
    Campinas, Brazil \\
    \texttt{pereira\_felipe@ic.unicamp.br}
    \And
    \href{https://orcid.org/0000-0002-9529-4253}{\includegraphics[scale=0.06]{orcid.pdf}\hspace{1mm}Pedro J. {de Rezende}}\\
    Institute of Computing\\
    University of Campinas\\
    Campinas, Brazil \\
    \texttt{pjr@unicamp.br}
    \And
    \href{https://orcid.org/0000-0002-8308-7812}{\includegraphics[scale=0.06]{orcid.pdf}\hspace{1mm}Tallys Yunes}\\
    Miami Herbert Business School\\
    University of Miami\\
    Coral Gables, USA \\
    \texttt{tallys@miami.edu}
    \And
    \href{https://orcid.org/0000-0003-0938-2941}{\includegraphics[scale=0.06]{orcid.pdf}\hspace{1mm}Luiz F. B. Morato}\\
    Institute of Computing\\
    University of Campinas\\
    Campinas, Brazil \\
    \texttt{fernandomorato93@gmail.com}
}

\renewcommand{\shorttitle}{Solving the Graph Burning Problem for Large Graphs}
\hypersetup{
pdftitle={Solving the Graph Burning Problem on Large Graphs},
pdfauthor={Felipe de C. Pereira, Pedro J. {de Rezende}, Tallys Yunes, Luiz F. B. Morato},
pdfkeywords={Graph burning, Burning number, Burning sequence, Set covering, Integer programming, Row generation},
}

\begin{document}

{
\thispagestyle{empty}
\addtocounter{page}{-1}
\hspace{0pt}
\vfill
\LARGE
\begin{center}
    \textbf{Notice}
\end{center}
\Large
This manuscript is a preliminary version of a conference paper presented at the 32nd Annual European Symposium on Algorithms (ESA 2024). The final version of this paper can be found at the following link: \href{https://doi.org/10.4230/LIPIcs.ESA.2024.94}{https://doi.org/10.4230/LIPIcs.ESA.2024.94}. We encourage readers to refer to the final paper for the most accurate and updated findings.
\vfill
\hspace{0pt}
}

\newpage

\maketitle

\begin{abstract}
We propose an exact algorithm for the Graph Burning Problem (\GBP), an \NPclass-hard optimization problem that models the spread of influence on social networks. Given a graph $G$ with vertex set $V$, the objective is to find a sequence of $k$ vertices in $V$, namely, $v_1, v_2, \dots, v_k$, such that $k$ is minimum and $\bigcup_{i = 1}^{k} \{u\! \in\! V\! : d(u, v_i) \leq k - i\} = V$, where $d(u,v)$ denotes the distance between $u$ and $v$. We formulate the problem as a set covering integer programming model and design a row generation algorithm for the \GBP. Our method exploits the fact that a very small number of covering constraints is often sufficient for solving the integer model, allowing the corresponding rows to be generated on demand. To date, the most efficient exact algorithm for the \GBP, denoted here by \GDCA, is able to obtain optimal solutions for graphs with up to 14,000 vertices within two hours of execution. In comparison, our algorithm finds provably optimal solutions approximately 236 times faster, on average, than \GDCA. For larger graphs, memory space becomes a limiting factor for \GDCA. Our algorithm, however, solves real-world instances with almost 200,000 vertices in less than 35 seconds, increasing the size of graphs for which optimal solutions are known by a factor of 14.
\end{abstract}


\section{Introduction}
\label{sec:introduction}

The Graph Burning Problem (\GBP) is a combinatorial optimization problem that models a form of contagion diffusion on social networks in which one seeks to propagate influence over the entire network as quickly as possible~\cite{Bonato2014}. By representing networks as graphs, a contagion can be thought of as a fire that spreads throughout the vertices of a graph following its adjacency relations.

In this problem, we are given an undirected graph $G = (V,E)$ that represents a social network, where each vertex in $V$ corresponds to an individual and each edge $\{u,v\} \in E$ indicates a reciprocal influence relationship between the individuals represented by $u$ and $v$.

A \textit{burning process} in $G$ unfolds in a series of \textit{rounds}. In every round $i \geq 1$, each vertex is either assigned the \textit{burned} or \textit{unburned} state. Initially, all vertices are unburned. 
In each round $i \geq 1$, exactly one vertex (a \textit{fire source}) is chosen to be set on fire and becomes burned. Moreover, starting in round $i = 2$, each unburned vertex that has at least one burned neighbor in round $i-1$ becomes burned and remains in that state until the last round.

A sequence $(v_1, v_2, \dots, v_k) \in V^k$, where $v_i$ is the $i$-th fire source, constitutes a \textit{burning sequence} for $G$, if the entire graph is burned by the $k$-th round.

For each $u \in V$, we denote by $N_j[u] = \{v \in V : d(u,v) \leq j\}$ the \textit{$j$-th closed neighborhood} of $u$, where $d(u,v)$ denotes the \textit{distance} between $u$ and $v$, \ie, the number of edges in a shortest path in $G$ that connects these vertices.

Formally, $(v_1, v_2, \dots, v_k)$ is a burning sequence for $G$ if
\begin{equation}
    \bigcup\nolimits_{i = 1}^{k} N_{k-i}[v_i] = V
    \label{eq:burning_sequence}
\end{equation}
It follows from the burning process that $v_i$ ascertains that each vertex in $N_{k - i}[v_i]$ (including $v_i$) gets burned by round $k$.

\begin{problem}[Graph Burning Problem]
Given an undirected graph $G$, find a burning sequence for $G$ of minimum length.
\end{problem}

The length of a shortest burning sequence for $G$ is called its \textit{burning number}, denoted $b(G)$, and was introduced in the literature as a graph parameter that measures the speed at which a propagation can spread throughout a network: the smaller the burning number of $G$ is, the more susceptible $G$ is to fast contagions. Although this parameter is of special interest for social networks, the problem has also been investigated for various other classes of graphs~\cite{Bessy2017,Bonato2021a,Bonato2021b,Bonato2019,Bonato2019b,Gupta2021,Liu2020,Liu2019,Mitsche2018,Sim2018,Tan2023}.

The \GBP is \NPclass-hard~\cite{Bessy2017,Bonato2014} and, so far, four exact approaches have been proposed to solve the problem for arbitrary graphs~\cite{Garcia2024,Garcia2022a}. The currently best known exact algorithm~\cite{Garcia2024}, referred to, here, as \GDCA, is able to find provably optimal solutions for real-world networks with up to 14,000 vertices in less than two hours. For larger graphs, memory space becomes a limiting factor.

\subsection{Our Contributions}
In this paper, we propose an exact algorithm for the \GBP based on an integer programming (\IP) formulation together with a row generation procedure.
Through a series of computational experiments, we are able to demonstrate that the proposed method significantly outperforms \GDCA. More specifically, our algorithm:
\begin{itemize}  
    \item Finds provably optimal solutions 236 times faster, on average, than \GDCA for networks with up to 14,000 vertices;
    \item Solves real-world networks with almost 200,000 vertices in less than 35 seconds, hence increasing the size of the vertex set of graphs for which optimal solutions are known by a factor of 14.
\end{itemize}

This paper is organized as follows. In Section~\ref{sec:previous_work}, we review the literature on the \GBP and, in Section~\ref{sec:algorithm}, we describe our exact algorithm. Section~\ref{sec:experiments} contains a report on computational experiments and an analysis of the results. Lastly, in Section~\ref{sec:concluding_remarks}, we present concluding remarks and address future work.

\section{Previous Work}
\label{sec:previous_work}

The \GBP was proposed in~\cite{Bonato2014} and has been extensively studied both from theoretical and practical points of view.
In this section, we present a brief background review of the problem for arbitrary graphs, focusing on upper bounds, heuristics, approximation algorithms, and mathematical models. For a probe regarding the \GBP on specific families of graphs, we refer the reader to the survey~\cite{Bonato2021a}.

Let $G = (V, E)$ be an arbitrary undirected graph with $p \geq 1$ connected components and let $n_1, n_2, \dots, n_p$ be the numbers of vertices of the these components. Denote by $b(G)$ the burning number of $G$.

Theorem~\ref{theor1} provides the best known upper bound for $b(G)$~\cite{Bastide2022,Garcia2022a}.
\begin{theorem}
    \label{theor1}
    $b(G) \leq p + \sum_{i = 1}^{p} \left\lceil (4n_i/3)^{1/2} \right\rceil$.
\end{theorem}

Conjecture~\ref{conj1} suggests a tighter upper bound for $b(G)$, but that result remains open since the problem was introduced \cite{Bonato2014,Garcia2022a}.
\begin{conjecture}
    \label{conj1}
    $b(G) \leq \sum_{i = 1}^{p} \left\lceil n_i^{1/2} \right\rceil$.
\end{conjecture}

Among the plethora of heuristics proposed for the \GBP~\cite{Farokh2020,Garcia2024,Gautam2022a,Nazeri2023,Simon2019b,Simon2019a}, a centrality-based genetic algorithm~\cite{Nazeri2023} and the greedy algorithm from~\cite{Garcia2024} are the latest in the literature.

Two 3-approximation algorithms for the \GBP were proposed in~\cite{Bessy2017,Bonato2019}, and more recently, a $\left(3 - 2/b(G)\right)$-approximation, referred to, here, as \BFF, was introduced in~\cite{Garcia2022b}. Algorithm~\ref{alg_bff} describes \BFF, which progressively builds a burning sequence $S$ by iteratively selecting the $i$-th fire source as the vertex that is farthest from any of the previous selected fire sources. In~\cite{Garcia2022b}, it is shown that $|S| \leq 3 \cdot b(G) - 2$ and that \BFF has worst-case time complexity $\bigO(|V|^2)$, provided that the distances between all pairs of vertices are computed a priori.

\begin{algorithm2e}[ht]
    \caption{\BFF (as used in~\cite{Garcia2024})}
    \label{alg_bff}
    \DontPrintSemicolon
    \SetKwInOut{Input}{Input}
    \SetKwInOut{Output}{Output}
    \Input{Graph $G = (V,E)$; distances between all pairs of vertices of $G$}
    \Output{A burning sequence $S$}
    Select $v_1$ arbitrarily; $S \leftarrow (v_1)$; $i \leftarrow 2$;\;
    \While{$S$ is not feasible}
    {
        $v_i = \underset{u \in V}{\arg\max}\;\left(\min\{d(u,v_1), d(u,v_2), \dots, d(u,v_{i-1}\right)\})$; \label{step:choosevi}\;
        $S \leftarrow (v_1, v_2, \dots, v_i)$; $i \leftarrow i + 1$;\;
    }
    \Return{$S$}
\end{algorithm2e}

\subsection{Existing Mathematical Models}

Regarding exact formulations for the \GBP, three \IP models, namely, \ILP, \CSP1 and \CSP2, were proposed in~\cite{Garcia2022a}. While \ILP consists of an optimization model, the last two are decision models that, for a given integer $B$, determine whether a burning sequence of length $B$ exists.

In both \ILP and \CSP1, the main variables are indexed by each pair $(v, i) \in V \times \{1, 2, \dots, U\}$, where $U$ is a known upper bound for $b(G)$, and each of them indicates whether $v$ is burned in round $i$. This idea has also been applied for the design of mathematical models for related problems, such as the well studied Target Set Selection Problem and some of its variants~\cite{Pereira2021thesis,Pereira2023,Pereira2021lagos,Shakarian2013}.

In \CSP2, the main variables are indexed by each pair $(u,v) \in V\times V$ and each of them specifies whether $u$ is responsible for $v$ getting burned, if $u$ is a fire source. The assemblage of \CSP2 requires that the distances between every pair of vertices be known.

\subsection{The Current Best Known Exact Algorithm}

We now describe \GDCA, an exact algorithm that leads to better performance results when compared to simply solving the models cataloged in the previous section, as was empirically demonstrated by experiments reported in~\cite{Garcia2024}.

\GDCA relies on the fact that the \GBP can be seen as a set covering problem. This was first observed in~\cite{Bonato2016} and later formalized in~\cite{Garcia2024} by means of a reduction of the \GBP to the Clustered Maximum Coverage Problem (\CMCP)~\cite{Garcia2024} that we now describe.

\begin{problem}[Clustered Maximum Coverage Problem]
    Given a set $P$ and $k \geq 1$ sets (clusters) $C_1, C_2, \dots, C_k$, each one containing subsets of $P$, find $k$ sets $S_1, S_2, \dots, S_k$ such that $S_i \in C_i$, for $i = 1, 2, \dots, k$, and $\left|\bigcup_{i = 1}^k S_i\right|$ is maximum.
\end{problem}

Given an undirected graph $G = (V,E)$, let $P = V$, $k = B$ and, for each $i = 1, 2, \dots, B$, take $C_i = \{S_{i,v} : v \in V\}$, where $S_{i,v} = N_{B-i}[v]$. The value of an optimal solution for \CMCP corresponds to the maximum number of vertices that can be burned in $G$ using a sequence of $B$ vertices~\cite{Garcia2024}. If such number equals $|V|$, then an optimal solution for \CMCP, say, $S_{1, v_1}, S_{2, v_2}, \dots, S_{k, v_k}$, corresponds to a burning sequence $(v_1, v_2, \dots, v_k)$ for $G$. Otherwise, one can conclude that $b(G) > B$.

In~\cite{Garcia2024}, the following \IP model, originally designed for the \CMCP, and referred to, here, as \CMCPIP, is used to decide whether a burning sequence of length $B$ for $G$ exists. Let $X = \{x_{v,i} : v \in V, i \in \{1, 2, \dots, B\}\}$ and $Y = \{y_v : v \in V\}$ be sets of binary variables such that $x_{v,i} = 1$ iff $v$ is the $i$-th fire source and $y_v = 1$ iff $v$ gets burned during the burning process.

\begin{empheq}[left = \CMCPIP\empheqlbrace\quad]{align}
    \max &\sum_{v \in V} y_v \label{cmcp_obj} \\
    \mbox{s.t.}\hspace{1.5cm}
    \sum_{v \in V} x_{v,i} &= 1 & \forall i \in \{1,2, \dots, B\} \label{cmcp_c1}\\
    \sum_{i = 1}^{B} \;\; \sum_{u \in V : v \in N_{B - i}[u]} x_{u,i} &\geq y_v &\forall v \in V \label{cmcp_c2}
\end{empheq}

The objective function~\eqref{cmcp_obj} maximizes the number of burned vertices. Constraints~\eqref{cmcp_c1} establish that exactly one vertex is assigned to each position in the burning sequence.  Lastly, Constraints~\eqref{cmcp_c2} ensure that if $v$ is in the burned state, then there is at least one fire source $u$ (possibly $v$ itself) such that $d(u,v) \leq B - i$, where $i$ is the position of $u$ in the burning sequence. \CMCPIP has a total of $\bigO(|V| \cdot B)$ binary variables and $\bigO(|V|)$ constraints. Loading this model requires that the distances between every pair of vertices be known.

\GDCA performs a binary search in a certain interval of candidate values for $b(G)$ and uses \CMCPIP to solve each of the decision problems encountered during the search. Algorithm~\ref{alg_garcia} describes the procedure.

\begin{algorithm2e}[ht]
    \caption{\GDCA}
    \label{alg_garcia}
    \DontPrintSemicolon
    \SetKwFunction{BFF}{BFF}
    \SetKwFunction{SolveCMCPIP}{SolveCMCP-IP}
    \SetKwInOut{Input}{Input}
    \SetKwInOut{Output}{Output}
    \Input{Graph $G = (V,E)$}
    \Output{Optimal burning sequence $S$}

    $D \leftarrow $ distance matrix of $G$; $S \leftarrow $ \BFF{$G, D$};\; $U \leftarrow |S|$; $L \leftarrow \left\lceil (|S| + 2)/3 \right\rceil$;\;
    
    \While{$L \leq U$}
    {
        $B \leftarrow \lfloor(L+U)/2\rfloor$; $(\mathit{obj}, S') \leftarrow $ \SolveCMCPIP{$G$, $D$, $B$};\;

        \leIf{$\mathit{obj} = |V|$}
        {
            $S \leftarrow S'$; $U \leftarrow B - 1$;
        }
        {
            $L \leftarrow B + 1$;
        }
    }
    \Return{$S$}\;
\end{algorithm2e}

First, the algorithm computes the distances between all pairs of vertices (\eg, by $|V|$ breadth-first searches, totaling $\bigO(|V|^2 + |V|\cdot|E|)$ time). Then, $\BFF$ is applied to obtain a burning sequence $S$ for $G$. Next, \GDCA computes upper and lower bounds for $b(G)$, namely, $U = |S|$ and $L = \left\lceil (|S| + 2)/3 \right\rceil$. The lower bound follows from the fact that, since $\BFF$ is a $(3-2/b(G))$-approximation, $|S| \leq (3 - 2/b(G))\cdot b(G)$ and, therefore $\left\lceil (|S| + 2)/3 \right\rceil \leq b(G)$. Then, a binary search is employed to solve $\bigO(\log (U - L + 1))$ \GBP decision problems.

In~\cite{Garcia2024}, it is shown that \GDCA is able to find optimal solutions for networks with up to 12,000 vertices. In the next section, we show how to improve \GDCA to design a more effective and efficient exact algorithm for the \GBP.

\section{A Row Generation Algorithm}
\label{sec:algorithm}

In this section, we introduce an exact algorithm for the \GBP, denoted by \PRYM, preceded by the presentation of some useful results and an \IP formulation for which a row generation method is employed in \PRYM.

Let $G = (V,E)$ be an undirected graph. Since any sequence containing all vertices of $V$ constitutes a trivial feasible solution for $G$, we have that $b(G) \leq |V|$. Moreover, there exists a burning sequence of length $k$ for $G$ for each integer $k$, with $b(G) \leq k \leq |V|$ since we may append (dummy) fire sources to any given burning sequence.

Furthermore, although the definition of \GBP does not forbid burning sequences that contain reoccurring vertices, it is easy to prove that for each $k \in \Z$, where $b(G) \leq k \leq |V|$, there exists a burning sequence of length $k$ for $G$ that does not contain repeated vertices.

We now propose an integer program for the decision version of the \GBP, denoted by \GBPIP. This model determines whether there exists a burning sequence of length $B \leq |V|$ for $G$. Let $X = \{x_{v,i} : v \in V, i \in \{1, 2, \dots, B\}\}$ be a set of binary variables such that $x_{v,i} = 1$ iff $v$ is the $i$-th fire source in a burning sequence for $G$.

\begin{empheq}[left = \GBPIP\empheqlbrace\quad]{align}
    \text{Find } X & \label{gbp_obj}\\ 
    \mbox{s.t.}\hspace{1.2cm}
    \sum_{i = 1}^{B} x_{v,i} &\leq 1 & \forall v \in V \label{gbp_c1}\\
    \sum_{v \in V} x_{v,i} &= 1 & \forall i \in \{1,2, \dots, B\} \label{gbp_c2}\\
    \sum_{i = 1}^{B} \;\; \sum_{u \in V : v \in N_{B - i}[u]} x_{u,i} &\geq 1 &\forall v \in V \label{gbp_c3}
\end{empheq}

Constraints~\eqref{gbp_c1} ensure that each $v \in V$ appears at most once in the burning sequence. Constraints~\eqref{gbp_c2} establish that exactly one vertex is assigned to each position in the burning sequence. Lastly, Constraints~\eqref{gbp_c3} ensure that each vertex is burned by the end of round $B$, \ie, that Equation~\eqref{eq:burning_sequence} is satisfied. From now on, we also refer to~\eqref{gbp_c3} as \textit{covering constraints}. Whenever~\eqref{gbp_c3} is satisfied for a vertex $v$, we say that $v$ is \textit{covered}, otherwise, $v$ is \textit{uncovered}.

We remark that although Constraints~\eqref{gbp_c1} are not necessary for the correctness of the model, they cut off integer solutions with repeated vertices, which reduces the search space.

\GBPIP has a total of $\bigO(B \cdot |V|)$ binary variables and $\bigO(|V|)$ constraints. Observe that \GBPIP can be obtained from \CMCPIP by: removing the objective function~\eqref{cmcp_obj}, adding Constraints~\eqref{gbp_c1}, and setting $y_v = 1$ for each $v \in V$.

\begin{proposition}
\label{proposition:feasible}
If $X = \{x_{v,i} : v \in V, i \in \{1, 2, \dots, B\}\}$ is a feasible solution for \GBPIP, then there exists a burning sequence $S$ of length $B$ for $G$ such that
for each $v \in V$, if $x_{v,i} = 1$, then $v$ is the $i$-th fire source of $S$.
\end{proposition}
\begin{proof}
    Let $X$ be a feasible solution for \GBPIP. By~\eqref{gbp_c2}, for each $i \in \{1, 2, \dots, B\}$, there exists exactly one vertex $v$ such that $x_{v, i} = 1$. Take a sequence $S = (v_1, v_2, \dots, v_B)$ such that $x_{v_i, i} = 1$. Since $X$ satisfies~\eqref{gbp_c3}, $S$ satisfies Equation~\eqref{eq:burning_sequence} and, therefore, $S$ is a burning sequence for $G$.
\end{proof}

\begin{proposition}
    \label{proposition:not_feasible}
    If there is a burning sequence of length $B$ for $G$, then \GBPIP is feasible.
\end{proposition}
\begin{proof}
    Let $S = (v_1, v_2, \dots, v_B)$ be a burning sequence for $G$ with no repeated vertices. Take $X = \{x_{v,i} : v \in V, i \in \{1, 2, \dots, B\}\}$ such that $x_{v,i} = 1$ iff $v$ is the $i$-th fire source of $S$. By construction, $X$ satisfies Constraints~\eqref{gbp_c1} and~\eqref{gbp_c2}. Also, since $S$ satisfies Equation~\eqref{eq:burning_sequence}, $X$ satisfies Constraints~\eqref{gbp_c3}. Therefore, $X$ is a feasible solution for \GBPIP.
\end{proof}

Now, let $L$ and $U$ be lower and upper bounds for $b(G)$. Since there is no burning sequence for $G$ of length less than $b(G)$, it follows from Proposition~\ref{proposition:feasible} that for every $B \in [L, b(G) - 1]$, \GBPIP is infeasible. Similarly, since there is a burning sequence for $G$ of length $b(G) + q$ for every $q \in \Z_{\geq 0}$, it follows from Proposition~\ref{proposition:not_feasible} that for every $B \in [b(G), U]$, \GBPIP admits a feasible solution from which a burning sequence of length $B$ can be built. Therefore, one can perform a binary search to determine the smallest value $B$ in the interval $[L, U]$ for which a feasible solution of \GBPIP exists, leading to an optimal solution of length $B$ for $G$. This idea is similar to the one employed in \GDCA (see Algorithm~\ref{alg_garcia}) and is the core of \PRYM, which is described in Algorithm~\ref{alg_felipe}. 

\begin{algorithm2e}[ht]
    \caption{\PRYM}
    \label{alg_felipe}
    \DontPrintSemicolon
    \SetKwFunction{BFFd}{BFF-d}
    \SetKwFunction{SolveGBPIP}{SolveGBP-IP}
    \SetKwInOut{Input}{Input}
    \SetKwInOut{Output}{Output}
    \Input{Graph $G = (V,E)$}
    \Output{Optimal burning sequence $S$}

    $S \leftarrow $ \BFFd{$G$}; $U \leftarrow |S|$; $L \leftarrow \left\lceil (|S| + 2)/3 \right\rceil$;\;
    
    \While{$L < U$}
    {
        $B \leftarrow \lfloor(L+U)/2\rfloor$; $(\mathit{answer}, S') \leftarrow $ \SolveGBPIP{$G$, $B$};\;

        \leIf{$\mathit{answer} = \mathit{feasible}$}
        {
            $S \leftarrow S'$; $U \leftarrow B$;
        }
        {
            $L \leftarrow B + 1$;
        }
    }
    \Return{$S$}\;
\end{algorithm2e}

First, \PRYM obtains a feasible solution $S$ by running a modified version of \BFF (\BFFd) which, instead of being provided with all pairwise distances between vertices, as in \GDCA, computes only the required distances on demand. In other words, during the $i$-th iteration of \BFFd, right before vertex $v_i$ is selected as the $i$-th fire source of $S$ (see step~\ref{step:choosevi} of Algorithm~\ref{alg_bff}), \BFFd computes the distance between $v_{i-1}$ and each vertex in $V$ by means of a single breadth-first search. These distances are stored in memory until \PRYM halts. The motivation for this change is that in that iteration only the distances from each vertex in $V$ to the $i-1$ previous fire sources are needed. Since the lengths of burning sequences are often much smaller than $|V|$, this modification speeds up \BFFd to a total complexity of $\bigO(|S|(|V| + |E|))$, leading to a significant improvement in practice.

Next, \PRYM computes lower and upper bounds for searching for $b(G)$, namely, $L = \left\lceil (|S| + 2)/3 \right\rceil$ and $U = |S|$. Lastly, \PRYM performs a binary search on the interval of values between $L$ and $U$, solving $\bigO(\log(U - L))$ \GBP decision problems by means of solving the \GBPIP model on each query.

We remark that to solve the \GBPIP model, \PRYM does not need to load the whole set of covering constraints from the start. Instead, these constraints are loaded on demand following the traditional \textit{lazy constraint} strategy: whenever the \IP solver finds an integer solution, we separate a violated covering constraint, if it exists, and add it to the model as a lazy constraint. This approach comes from the observation that, very often, a small subset of the covering constraints may be sufficient to prove that \GBPIP is either infeasible or feasible.

\begin{figure}[ht]
    \begin{subfigure}[ht]{0.325\textwidth}
        \centering
        \includegraphics[width=0.65\textwidth]{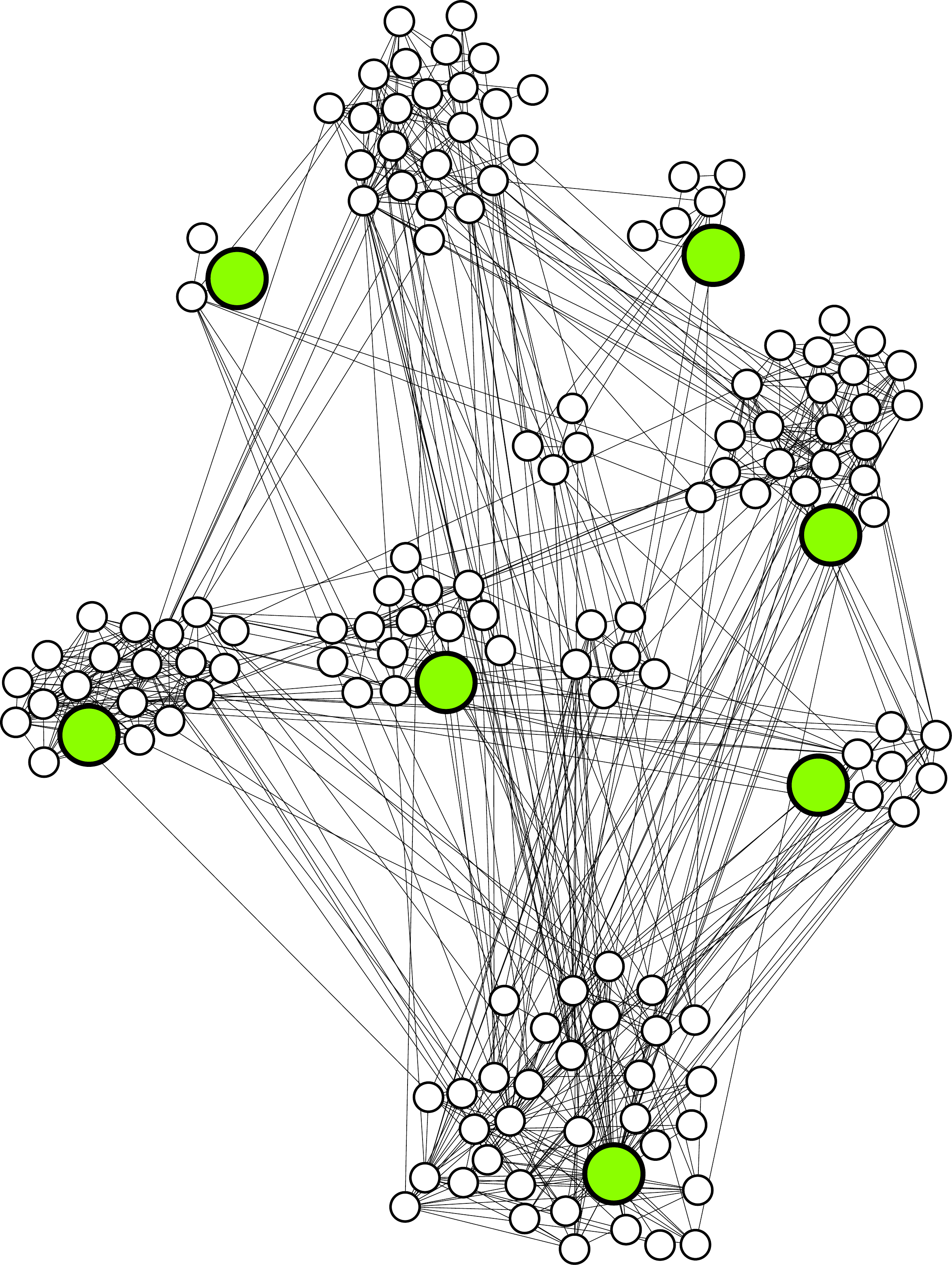}
        \caption{\tt ia-enron-only}
        \label{fig:ia-enron-only}
    \end{subfigure}
    \hfill
    \begin{subfigure}[ht]{0.325\textwidth}
        \centering
        \includegraphics[width=0.8\textwidth]{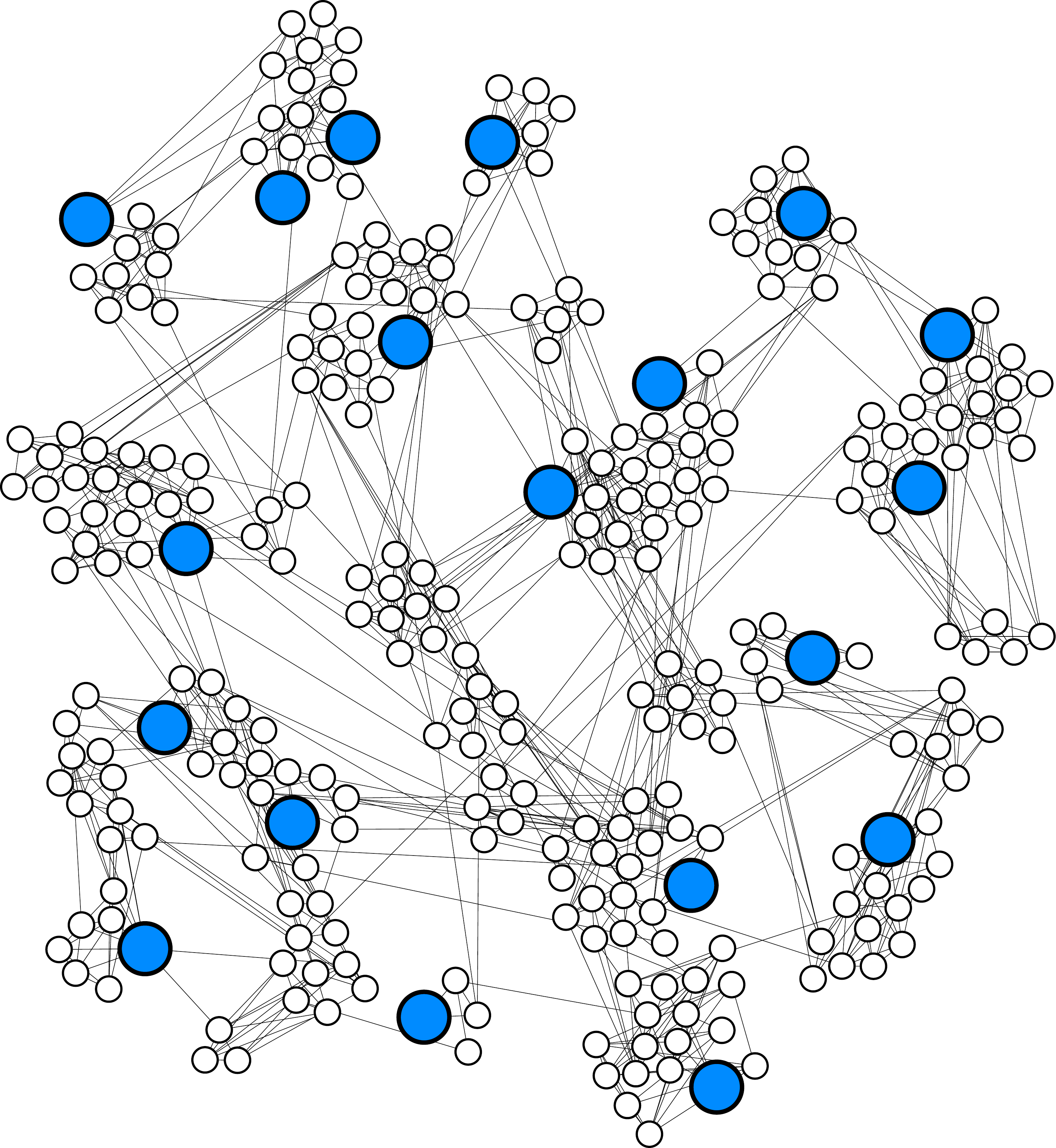}
        \caption{\tt DD244}
        \label{fig:dd244}
    \end{subfigure}
    \hfill
    \begin{subfigure}[ht]{0.325\textwidth}
        \centering
        \includegraphics[width=0.95\textwidth]{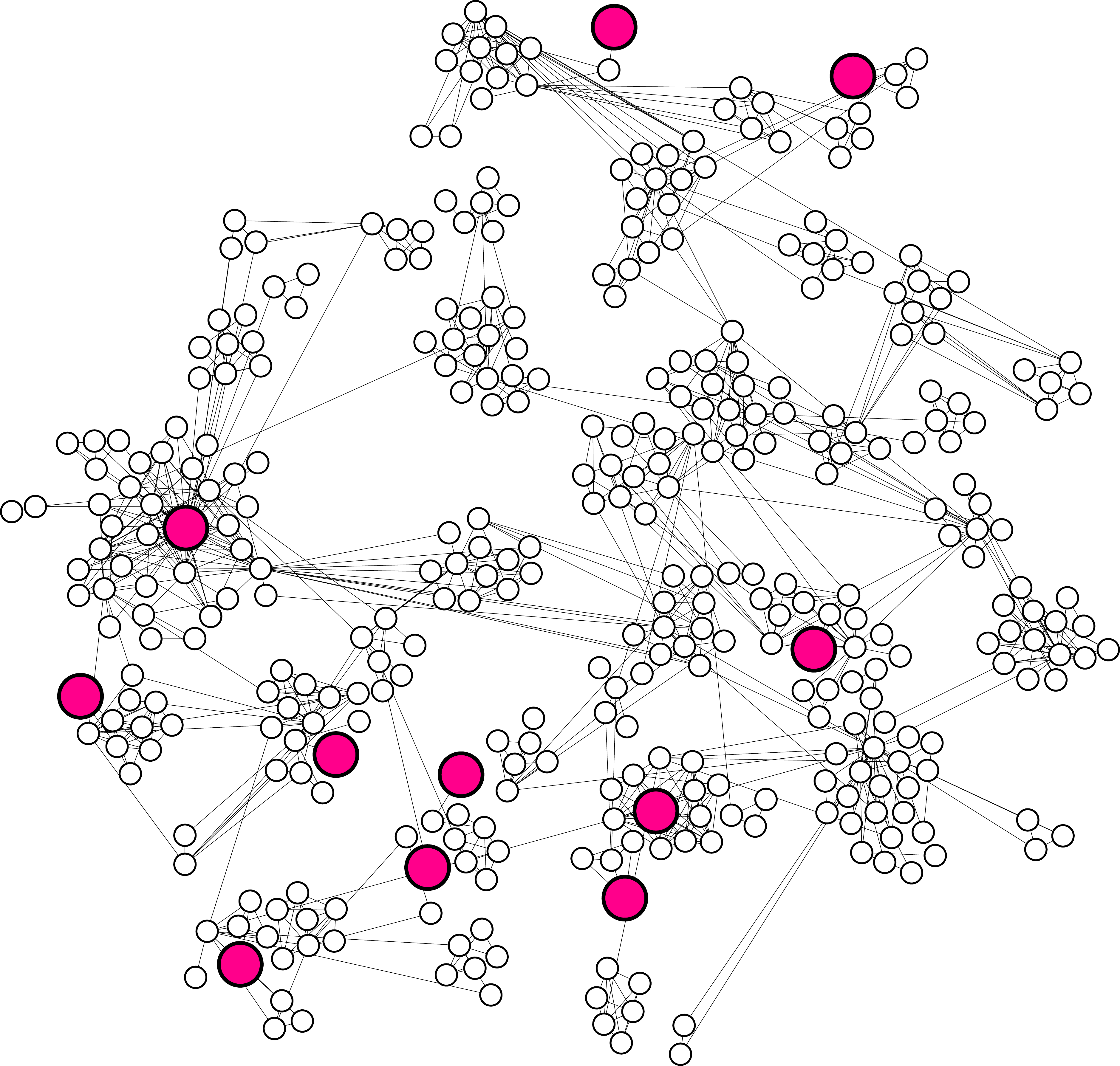}
        \caption{\tt ca-netscience}
        \label{fig:ca-netscience}
    \end{subfigure}
    \caption{Illustration of the vertices (colored and enlarged) whose covering constraints suffice to prove the infeasibility of \GBPIP for $B = b(G) - 1$.} \label{fig:covering_example}
\end{figure}

As an illustration, consider these real-world networks obtained from \cite{network_reposity}: {\tt ia-enron-only} ($|V| = 143$, $|E| = 623$, $b(G) = 4$); {\tt DD244} ($|V| = 291$, $|E| = 822$, $b(G) = 7$); and {\tt ca-netscience} ($|V| = 379$, $|E| = 914$, $b(G) = 6$) and depicted in Figure~\ref{fig:covering_example}.
For each of them, the infeasibility of the \GBPIP model for $B = b(G) - 1$ can be established by loading only the covering constraints associated with the colored vertices. Moreover, in Section~\ref{sec:experiments}, we show that, for instances containing hundreds of thousands vertices, just dozens covering constraints are sufficient to prove that there is no burning sequence of length $B = b(G) - 1$.

In light of that, whenever \PRYM invokes an \IP solver to solve \GBPIP, only the covering constraints associated to the vertices in the burning sequence obtained by \BFFd are initially loaded into the model. The rest of the covering constraints used by the solver are added according to the following separation procedure.

Given an integer solution found by the solver, we first extract the sequence $S = (v_1, v_2, \dots, v_B)$ such that $x_{v_i, i } = 1$ in $\bigO(|V| \cdot B)$ time. Then, for each $i \in \{1, 2, \dots, B\}$, we determine the distances between $v_i$ and all vertices of $G$ by breadth-first search, in $\bigO(|V| + |E|)$ time, whenever they had not been previously computed by \PRYM. Next, for each $u \in V$,  we calculate the distance between $u$ and its closest vertex among those burned by round $B$ in the burning process. This value, denoted by $d_S(u)$, can be calculated in $\bigO(B)$ time as
$$d_S(u) = \max(0, \min\{d(u, v_1) - (B - 1), d(u, v_2) - (B - 2), \dots, d(u,v_B)\}).$$

If $d_S(u) = 0$, then $u$ is covered, otherwise, $d_S(u) \geq 1$ and $u$ is uncovered. If $d_S(v) = 0$ for every $v \in V$, then no constraint is violated and $S$ is a burning sequence for $G$. Otherwise, we select $w = {\arg\max}_{u \in V}\; d_S(u)$, \ie, the uncovered vertex that is farthest from any burned vertex. Then, we calculate the distances between $w$ and all vertices of $G$ by a breadth-first search, in $\bigO(|V| + |E|)$ time, if they were not calculated previously. Lastly, we compute the covering constraint for $w$ and load it onto the solver as a lazy constraint.

As a branching rule for solving the \GBPIP, \PRYM determines that for each vertex $v$, the variable $x_{v,i}$ has a higher priority for branching than $x_{v, j}$ for every $j > i$. The purpose of this approach is to decide the first positions of the burning sequence earlier in the search.

We now highlight the advantages of \PRYM over \GDCA. First, observe that \PRYM addresses the \GBP decision problem directly via the \GBPIP model, while \GDCA attempts to solve, via the \CMCPIP model, an instance of an optimization problem all the way through, even when $B < b(G)$ and an upper bound less than $|V|$ for the objective function~\eqref{cmcp_obj} is determined. This shows that \PRYM can be particularly more expedient than \GDCA in these cases by saving valuable computing time.

Another major time-saving strategy is that \PRYM computes, on demand, only distances between pairs of vertices of $G$ known to be necessary instead of \GDCA's $|V|^2$ such computations. As we show in Section~\ref{sec:experiments}, memory space becomes a limiting factor for \GDCA for graphs with upwards of 14,000 vertices, while \PRYM is able to handle instances with hundreds of thousands of vertices.

Lastly, a crucial advantage stems from the row generation approach whose separation algorithm is able to discover a small number of decisive covering constraints that are often sufficient for the solver to find a feasible solution or to prove the infeasibility of \GBPIP for a given value of the parameter $B$. Moreover, since covering constraints tend to involve a substantial number of variables for large graphs, loading a small subset of these constraints ultimately speeds up the resolution of the linear relaxation.

In the next section, we report a series of experiments comparing the efficiency and efficacy of \PRYM and \GDCA on a large benchmark of instances.

\section{Computational Experiments}
\label{sec:experiments}

We now describe the experiments we carried out to empirically evaluate \PRYM. For this purpose, we used a machine equipped with an Intel\textsuperscript{\textregistered} Xeon\textsuperscript{\textregistered} E5-2630 v4 processor, 64 GB of RAM, and the Ubuntu 22.04.1 LTS operating system. For \IP solver, we used Gurobi v10.0.3 running on a single thread of execution.
The benchmark of instances employed here extends the one used in the experiments reported in~\cite{Garcia2024} to a total of 66 real-world networks obtained from the Network Repository~\cite{network_reposity} and the Stanford Large Network Dataset Collection~\cite{snap}.

The instances were divided into two sets according to the number of vertices: $\Delta_{\mathrm{10K}}$ is the set of instances with at most 10K vertices and $\Delta_{\mathrm{200K}}$ comprises the instances with more than 10K vertices and at most 200K vertices.
We refer the reader to our publicly available 
repository~\cite{complementary-data} where the source code, instances, and solutions obtained are accessible.

The results for $\Delta_{\mathrm{10K}}$ and $\Delta_{\mathrm{200K}}$ are presented in Tables~\ref{tab_delta_10K}, \ref{tab_delta_200K}, respectively.
The first three columns describe the instances while the two subsequent ones display the lower and upper bounds for $b(G)$, $L$ and $U$, obtained via \BFFd. The ``Opt'' column shows the optimal value, $b(G)$, while the following two columns display the running times (rounded up to the nearest second) spent by \GDCA and \PRYM to find provably optimal solutions.
The last two columns show the number of covering constraints (of type~\eqref{gbp_c3}) loaded by \PRYM while solving \GBPIP for $B = b(G) - 1$ and for $B = b(G)$.

Recall that when $B = b(G) - 1$, \GBPIP is infeasible and $B$ is a lower bound for $b(G)$.
On the other hand, when $B = b(G)$, \GBPIP admits a feasible solution and $B$ is an upper bound for $b(G)$.
We highlight that according to Tables~\ref{tab_delta_10K} and \ref{tab_delta_200K}, the number of covering constraints used by \PRYM for these cases falls significantly short of the number of vertices in each graph. In fact, when $B = b(G) - 1$ and $B = b(G)$, \PRYM had to load, on average, only $5.54\%$ and $7.14\%$ of the whole set of possible covering constraints, respectively.

Also, the burning numbers that appear in bold in Tables~\ref{tab_delta_200K} are newly proven optimal results. In the running time columns, the entries containing `--' indicate that the execution was halted due to memory overflow. Also, in the last column, whenever the initial upper bound $U$ was equal to the burning number, `--' is used to reflect that there was no need for \PRYM to solve \GBPIP for $B = U = b(G)$.

Considering the 50 instances, 48 from $\Delta_{\mathrm{10K}}$ and 2 from $\Delta_{\mathrm{200K}}$, for which both \PRYM and \GDCA obtained provably optimal solutions, for 27 instances \PRYM was on average 236.7 times faster than \GDCA; they attained the same running times for 21 instances, considering the granularity of the time measurements presented (\textrm{sec});
and \PRYM only performed slower than \GDCA by a few seconds on two instances: {\tt DD349} and {\tt bal-bin-tree-9}. Among these 50 instances, the most remarkable result occurred for the {\tt athletes} instance, which was solved by \PRYM in 2 seconds, in contrast to the 5263 seconds spent by \GDCA.

The remaining 16 instances from $\Delta_{\mathrm{200K}}$ were only solved by \PRYM, which spent a maximum of 34 seconds per instance. Remarkably, despite the large sizes of these graphs, which have up to 200,000 vertices, \PRYM required just a few dozen covering constraints to solve \GBPIP.

\begin{table}[ht]
\rowcolors{13}{gray!30}{white}
\centering
\small
\caption{Quantifying the empirical results obtained for $\Delta_{\mathrm{10K}}$.}
\label{tab_delta_10K}
\begin{tabular}{lrrlrrlrlrrlrr}
\multicolumn{3}{c}{Instance} & ~ &
\multicolumn{2}{c}{Bounds} & ~ &
\multicolumn{1}{c}{Opt} & ~ &
\multicolumn{2}{c}{Time (sec)} & ~ &
\multicolumn{2}{c}{\#Cov. Constr. for:}\\
\cmidrule{1-3} \cmidrule{5-6} \cmidrule{8-8} \cmidrule{10-11} \cmidrule{13-14}
Name &
  \multicolumn{1}{r}{$|V|$} &
  \multicolumn{1}{r}{$|E|$} &
   &
  \multicolumn{1}{r}{$L$} &
  \multicolumn{1}{r}{$U$} &
   &
   $b(G)$
   &
   &
  \multicolumn{1}{r}{\GDCA} &
  \multicolumn{1}{r}{\PRYM} &
  &
  \multicolumn{1}{r}{$b(G)-1$} &
  \multicolumn{1}{r}{$b(G)$} \\ \cmidrule{1-3} \cmidrule{5-6} \cmidrule{8-8} \cmidrule{10-11} \cmidrule{13-14}
karate           & 34   & 78     & ~ & 2 & 4  & ~ & 3  & ~ & 1    & 1    & ~ & 4   & 7  \\
chesapeake       & 39   & 170    & ~ & 2 & 3  & ~ & 3  & ~ & 1    & 1    & ~ & 7   & -- \\
dolphins         & 62   & 159    & ~ & 3 & 6  & ~ & 4  & ~ & 1    & 1    & ~ & 7   & 12 \\
rt-retweet       & 96   & 117    & ~ & 3 & 5  & ~ & 5  & ~ & 1    & 1    & ~ & 9   & -- \\
polbooks         & 105  & 441    & ~ & 3 & 5  & ~ & 4  & ~ & 1    & 1    & ~ & 10  & 9  \\
adjnoun          & 112  & 425    & ~ & 2 & 4  & ~ & 4  & ~ & 1    & 1    & ~ & 6   & -- \\
ia-infect-hyper  & 113  & 2196   & ~ & 2 & 3  & ~ & 3  & ~ & 1    & 1    & ~ & 9   & -- \\
C125-9           & 125  & 6963   & ~ & 2 & 3  & ~ & 3  & ~ & 1    & 1    & ~ & 42  & -- \\
ia-enron-only    & 143  & 623    & ~ & 3 & 5  & ~ & 4  & ~ & 1    & 1    & ~ & 7   & 12 \\
c-fat200-1       & 200  & 1534   & ~ & 3 & 7  & ~ & 7  & ~ & 1    & 1    & ~ & 37  & -- \\
c-fat200-2       & 200  & 3235   & ~ & 3 & 5  & ~ & 5  & ~ & 1    & 1    & ~ & 16  & -- \\
c-fat200-5       & 200  & 8473   & ~ & 2 & 3  & ~ & 3  & ~ & 1    & 1    & ~ & 5   & -- \\
sphere           & 258  & 1026   & ~ & 4 & 9  & ~ & 7  & ~ & 2    & 1    & ~ & 44  & 33 \\
DD244            & 291  & 822    & ~ & 4 & 10 & ~ & 7  & ~ & 1    & 1    & ~ & 19  & 41 \\
ca-netscience    & 379  & 914    & ~ & 4 & 8  & ~ & 6  & ~ & 1    & 1    & ~ & 11  & 25 \\
infect-dublin    & 410  & 2765   & ~ & 3 & 6  & ~ & 5  & ~ & 1    & 1    & ~ & 12  & 8  \\
c-fat500-1       & 500  & 4459   & ~ & 5 & 11 & ~ & 9  & ~ & 1    & 1    & ~ & 33  & 72 \\
c-fat500-2       & 500  & 9139   & ~ & 4 & 8  & ~ & 7  & ~ & 1    & 1    & ~ & 36  & 13 \\
c-fat500-5       & 500  & 23191  & ~ & 3 & 5  & ~ & 5  & ~ & 1    & 1    & ~ & 32  & -- \\
bio-diseasome    & 516  & 1188   & ~ & 4 & 10 & ~ & 7  & ~ & 1    & 1    & ~ & 13  & 15 \\
web-polblogs     & 643  & 2280   & ~ & 3 & 7  & ~ & 5  & ~ & 2    & 1    & ~ & 7   & 15 \\
DD687            & 725  & 2600   & ~ & 4 & 9  & ~ & 7  & ~ & 21   & 2    & ~ & 17  & 85 \\
rt-twitter-copen & 761  & 1029   & ~ & 4 & 8  & ~ & 7  & ~ & 2    & 1    & ~ & 9   & 13 \\
DD68             & 775  & 2093   & ~ & 5 & 12 & ~ & 9  & ~ & 7    & 2    & ~ & 28  & 85 \\
ia-crime-moreno  & 829  & 1475   & ~ & 3 & 7  & ~ & 7  & ~ & 21   & 1    & ~ & 58  & -- \\
DD199            & 841  & 1902   & ~ & 7 & 18 & ~ & 12 & ~ & 9    & 6    & ~ & 96  & 87 \\
soc-wiki-Vote    & 889  & 2914   & ~ & 3 & 6  & ~ & 6  & ~ & 4    & 1    & ~ & 15  & -- \\
DD349            & 897  & 2087   & ~ & 6 & 16 & ~ & 12 & ~ & 7    & 13   & ~ & 113 & 110\\
DD497            & 903  & 2453   & ~ & 6 & 15 & ~ & 10 & ~ & 14   & 14   & ~ & 30  & 146\\
socfb-Reed98     & 962  & 18812  & ~ & 2 & 4  & ~ & 4  & ~ & 4    & 1    & ~ & 7   & -- \\
lattice3D        & 1000 & 2700   & ~ & 5 & 12 & ~ & 10 & ~ & 1767 & 1091 & ~ & 295 & 103\\
bal-bin-tree-9   & 1023 & 1022   & ~ & 5 & 11 & ~ & 10 & ~ & 1    & 7    & ~ & 513 & 16 \\
delaunay-n10     & 1024 & 3056   & ~ & 5 & 11 & ~ & 9  & ~ & 42   & 9    & ~ & 58  & 140\\
stufe            & 1036 & 1868   & ~ & 6 & 15 & ~ & 12 & ~ & 424  & 147  & ~ & 204 & 163\\
lattice2D        & 1089 & 2112   & ~ & 7 & 19 & ~ & 13 & ~ & 1424 & 95   & ~ & 107 & 264\\
bal-ter-tree-6   & 1093 & 1092   & ~ & 4 & 8  & ~ & 7  & ~ & 1    & 1    & ~ & 44  & 21 \\
email-univ       & 1133 & 5451   & ~ & 3 & 6  & ~ & 5  & ~ & 7    & 1    & ~ & 9   & 18 \\
econ-mahindas    & 1258 & 7513   & ~ & 3 & 6  & ~ & 5  & ~ & 32   & 1    & ~ & 10  & 15 \\
ia-fb-messages   & 1266 & 6451   & ~ & 3 & 5  & ~ & 5  & ~ & 9    & 1    & ~ & 9   & -- \\
bio-yeast        & 1458 & 1948   & ~ & 5 & 11 & ~ & 9  & ~ & 23   & 1    & ~ & 14  & 27 \\
tech-routers-rf  & 2113 & 6632   & ~ & 3 & 7  & ~ & 6  & ~ & 30   & 1    & ~ & 9   & 21 \\
chameleon        & 2277 & 36101  & ~ & 3 & 6  & ~ & 6  & ~ & 36   & 1    & ~ & 12  & -- \\
tvshow           & 3892 & 17262  & ~ & 4 & 10 & ~ & 9  & ~ & 302  & 2    & ~ & 14  & 18 \\
facebook         & 4039 & 88234  & ~ & 3 & 5  & ~ & 4  & ~ & 26   & 1    & ~ & 5   & 5  \\
DD6              & 4152 & 10320  & ~ & 9 & 24 & ~ & 16 & ~ & 716  & 189  & ~ & 135 & 221\\
squirrel         & 5201 & 198493 & ~ & 3 & 6  & ~ & 6  & ~ & 405  & 1    & ~ & 9   & -- \\
politician       & 5908 & 41729  & ~ & 4 & 8  & ~ & 7  & ~ & 452  & 2    & ~ & 11  & 14 \\
government       & 7057 & 89455  & ~ & 3 & 6  & ~ & 6  & ~ & 749  & 1    & ~ & 10  & -- \\\hline
\end{tabular}
\end{table}

\clearpage

\begin{table}[ht]
\rowcolors{13}{gray!30}{white}
\centering
\small
\caption{Quantifying the empirical results obtained for $\Delta_{\mathrm{200K}}$. Newly established burning numbers appear in bold.}
\label{tab_delta_200K}
\begin{tabular}{lrrlrrlrlrrlrr}
\multicolumn{3}{c}{Instance} & ~ &
\multicolumn{2}{c}{Bounds} & ~ &
\multicolumn{1}{c}{Opt} & ~ &
\multicolumn{2}{c}{Time (sec)} & ~ &
\multicolumn{2}{c}{\#Cov. Constr. for:}\\
\cmidrule{1-3} \cmidrule{5-6} \cmidrule{8-8} \cmidrule{10-11} \cmidrule{13-14}
Name &
  \multicolumn{1}{r}{$|V|$} &
  \multicolumn{1}{r}{$|E|$} &
   &
  \multicolumn{1}{r}{$L$} &
  \multicolumn{1}{r}{$U$} &
   &
   $b(G)$
   &
   &
  \multicolumn{1}{r}{\GDCA} &
  \multicolumn{1}{r}{\PRYM} &
  &
  \multicolumn{1}{r}{$b(G)-1$} &
  \multicolumn{1}{r}{$b(G)$} \\ \cmidrule{1-3} \cmidrule{5-6} \cmidrule{8-8} \cmidrule{10-11} \cmidrule{13-14}
crocodile        & 11631 & 170918  & ~ & 3 & 6  & ~ & 6           & ~ & 1982 & 1  & ~ & 10 & -- \\
athletes         & 13866 & 86858   & ~ & 3 & 7  & ~ & \textbf{7}  & ~ & 5263 & 2  & ~ & 19 & -- \\
company          & 14113 & 52310   & ~ & 4 & 9  & ~ & \textbf{9}  & ~ & --   & 6  & ~ & 21 & -- \\
musae-facebook   & 22470 & 171002  & ~ & 4 & 9  & ~ & \textbf{8}  & ~ & --   & 7  & ~ & 13 & 22 \\
new-sites        & 27917 & 206259  & ~ & 4 & 8  & ~ & \textbf{8}  & ~ & --   & 4  & ~ & 15 & -- \\
deezer-europe    & 28281 & 92752   & ~ & 5 & 12 & ~ & \textbf{10} & ~ & --   & 24 & ~ & 16 & 18 \\
RO-gemsec-deezer & 41773 & 125826  & ~ & 4 & 10 & ~ & \textbf{10} & ~ & --   & 10 & ~ & 16 & -- \\
HU-gemsec-deezer & 47538 & 222887  & ~ & 4 & 9  & ~ & \textbf{8}  & ~ & --   & 30 & ~ & 14 & 39 \\
artist           & 50515 & 819306  & ~ & 3 & 7  & ~ & \textbf{6}  & ~ & --   & 10 & ~ & 8  & 11 \\
HR-gemsec-deezer & 54573 & 498202  & ~ & 3 & 7  & ~ & \textbf{7}  & ~ & --   & 8  & ~ & 12 & -- \\
soc-brightkite   & 56739 & 212945  & ~ & 4 & 9  & ~ & \textbf{9}  & ~ & --   & 17 & ~ & 15 & -- \\
socfb-OR         & 63392 & 816886  & ~ & 4 & 8  & ~ & \textbf{8}  & ~ & --   & 11 & ~ & 13 & -- \\
soc-slashdot     & 70068 & 358647  & ~ & 3 & 7  & ~ & \textbf{7}  & ~ & --   & 8  & ~ & 9  & -- \\
soc-BlogCatalog  & 88784 & 2093195 & ~ & 3 & 5  & ~ & \textbf{5}  & ~ & --   & 4  & ~ & 9  & -- \\
soc-buzznet         & 101163  & 2763066  & ~ & 2 & 4  & ~ & \textbf{4}  & ~ & -- & 21   & ~ & 40 & -- \\
soc-LiveMocha       & 104103  & 2193083  & ~ & 3 & 5  & ~ & \textbf{5}  & ~ & -- & 23   & ~ & 29 & -- \\
soc-douban          & 154908  & 327162   & ~ & 3 & 6  & ~ & \textbf{6}  & ~ & -- & 34   & ~ & 30 & -- \\
soc-gowalla         & 196591  & 950327   & ~ & 4 & 8  & ~ & \textbf{8}  & ~ & -- & 31   & ~ & 10 & -- \\\hline
\end{tabular}
\end{table}

\section{Concluding Remarks and Future Work}
\label{sec:concluding_remarks}

In this paper, we propose an exact algorithm for the \GBP, namely \PRYM. For a given arbitrary graph, \PRYM finds an optimal burning sequence by means of solving multiple decision problems formulated as set covering \IP models, while it generates essential covering constraints on demand. The algorithm takes advantage of the fact that a very small number of covering constraints is often sufficient for solving those decision problems, as was confirmed in practice.
Via computational experiments, we demonstrate that \PRYM far outperforms the previously best known exact algorithm for \GBP, and it is able to solve real-world instances with up to 200,000 vertices in less than 35 seconds. Results for even larger instances are forthcoming.

As for future research, we intend to investigate whether a column generation approach can also be successfully applied to the proposed \IP formulation, extending \PRYM to a branch-and-price algorithm and potentially increasing the suitability of the algorithm for solving the \GBP for larger graphs.

\bibliographystyle{plainurl}
\bibliography{references}  

\begin{thebibliography}{10}

\bibitem{Bastide2022}
Paul Bastide, Marthe Bonamy, Anthony Bonato, Pierre Charbit, Shahin Kamali,
  Th{\'e}o Pierron, and Mika{\"e}l Rabie.
\newblock Improved pyrotechnics: Closer to the burning number conjecture.
\newblock {\em The Electronic Journal of Combinatorics}, 30(4), 2023.
\newblock \href {https://doi.org/10.37236/11113} {\path{doi:10.37236/11113}}.

\bibitem{Bessy2017}
Stéphane Bessy, Anthony Bonato, Jeannette Janssen, Dieter Rautenbach, and
  Elham Roshanbin.
\newblock Burning a graph is hard.
\newblock {\em Discrete Applied Mathematics}, 232:73--87, 2017.
\newblock \href {https://doi.org/10.1016/j.dam.2017.07.016}
  {\path{doi:10.1016/j.dam.2017.07.016}}.

\bibitem{Bonato2021a}
Anthony Bonato.
\newblock A survey of graph burning.
\newblock {\em Contributions to Discrete Mathematics}, 16(1):185 – 197, 2021.
\newblock \href {https://doi.org/10.11575/cdm.v16i1.71194}
  {\path{doi:10.11575/cdm.v16i1.71194}}.

\bibitem{Bonato2021b}
Anthony Bonato, Sean English, Bill Kay, and Daniel Moghbel.
\newblock Improved bounds for burning fence graphs.
\newblock {\em Graphs and Combinatorics}, 37(6):2761--2773, 2021.
\newblock \href {https://doi.org/10.1007/s00373-021-02390-x}
  {\path{doi:10.1007/s00373-021-02390-x}}.

\bibitem{Bonato2014}
Anthony Bonato, Jeannette Janssen, and Elham Roshanbin.
\newblock Burning a graph as a model of social contagion.
\newblock In {\em Algorithms and Models for the Web Graph}, pages 13--22, 2014.
\newblock \href {https://doi.org/10.1007/978-3-319-13123-8_2}
  {\path{doi:10.1007/978-3-319-13123-8_2}}.

\bibitem{Bonato2016}
Anthony Bonato, Jeannette Janssen, and Elham Roshanbin.
\newblock How to burn a graph.
\newblock {\em Internet Mathematics}, 12(1-2):85--100, 2016.
\newblock \href {https://doi.org/10.1080/15427951.2015.1103339}
  {\path{doi:10.1080/15427951.2015.1103339}}.

\bibitem{Bonato2019}
Anthony Bonato and Shahin Kamali.
\newblock Approximation algorithms for graph burning.
\newblock In {\em Theory and Applications of Models of Computation}, pages
  74--92, 2019.
\newblock \href {https://doi.org/10.1007/978-3-030-14812-6_6}
  {\path{doi:10.1007/978-3-030-14812-6_6}}.

\bibitem{Bonato2019b}
Anthony Bonato and Thomas Lidbetter.
\newblock Bounds on the burning numbers of spiders and path-forests.
\newblock {\em Theoretical Computer Science}, 794:12--19, 2019.
\newblock \href {https://doi.org/https://doi.org/10.1016/j.tcs.2018.05.035}
  {\path{doi:https://doi.org/10.1016/j.tcs.2018.05.035}}.

\bibitem{Farokh2020}
Zahra~Rezai Farokh, Maryam Tahmasbi, Zahra Haj Rajab~Ali Tehrani, and Yousof
  Buali.
\newblock New heuristics for burning graphs, 2020.
\newblock \href {https://doi.org/10.48550/arXiv.2003.09314}
  {\path{doi:10.48550/arXiv.2003.09314}}.

\bibitem{Garcia2024}
Jesús García-Díaz and José~Alejandro Cornejo-Acosta.
\newblock A greedy heuristic for graph burning, 2024.
\newblock \href {https://doi.org/10.48550/arXiv.2401.07577}
  {\path{doi:10.48550/arXiv.2401.07577}}.

\bibitem{Garcia2022b}
Jesús García-Díaz, Julio~César Pérez-Sansalvador, Lil María~Xibai
  Rodríguez-Henríquez, and José~Alejandro Cornejo-Acosta.
\newblock Burning graphs through farthest-first traversal.
\newblock {\em IEEE Access}, 10:30395--30404, 2022.
\newblock \href {https://doi.org/10.1109/ACCESS.2022.3159695}
  {\path{doi:10.1109/ACCESS.2022.3159695}}.

\bibitem{Garcia2022a}
Jesús García-Díaz, Lil María~Xibai Rodríguez-Henríquez, Julio~César
  Pérez-Sansalvador, and Saúl~Eduardo Pomares-Hernández.
\newblock Graph burning: Mathematical formulations and optimal solutions.
\newblock {\em Mathematics}, 10(15), 2022.
\newblock \href {https://doi.org/10.3390/math10152777}
  {\path{doi:10.3390/math10152777}}.

\bibitem{Gautam2022a}
Rahul~Kumar Gautam, Anjeneya~Swami Kare, and Durga~Bhavani S.
\newblock Faster heuristics for graph burning.
\newblock {\em Applied Intelligence}, 52(2):1351--1361, 2022.
\newblock \href {https://doi.org/10.1007/s10489-021-02411-5}
  {\path{doi:10.1007/s10489-021-02411-5}}.

\bibitem{Gupta2021}
Arya~Tanmay Gupta, Swapnil~A. Lokhande, and Kaushik Mondal.
\newblock Burning grids and intervals.
\newblock In {\em Algorithms and Discrete Applied Mathematics}, pages 66--79,
  2021.
\newblock \href {https://doi.org/10.1007/978-3-030-67899-9_6}
  {\path{doi:10.1007/978-3-030-67899-9_6}}.

\bibitem{snap}
Jure Leskovec and Andrej Krevl.
\newblock {SNAP Datasets}: {Stanford} large network dataset collection.
\newblock \url{http://snap.stanford.edu/data}, June 2014.

\bibitem{Liu2020}
Huiqing Liu, Xuejiao Hu, and Xiaolan Hu.
\newblock Burning number of caterpillars.
\newblock {\em Discrete Applied Mathematics}, 284:332--340, 2020.
\newblock \href {https://doi.org/https://doi.org/10.1016/j.dam.2020.03.062}
  {\path{doi:https://doi.org/10.1016/j.dam.2020.03.062}}.

\bibitem{Liu2019}
Huiqing Liu, Ruiting Zhang, and Xiaolan Hu.
\newblock Burning number of theta graphs.
\newblock {\em Applied Mathematics and Computation}, 361:246--257, 2019.
\newblock \href {https://doi.org/https://doi.org/10.1016/j.amc.2019.05.031}
  {\path{doi:https://doi.org/10.1016/j.amc.2019.05.031}}.

\bibitem{Mitsche2018}
Dieter Mitsche, Paweł Prałat, and Elham Roshanbin.
\newblock Burning number of graph products.
\newblock {\em Theoretical Computer Science}, 746:124--135, 2018.
\newblock \href {https://doi.org/https://doi.org/10.1016/j.tcs.2018.06.036}
  {\path{doi:https://doi.org/10.1016/j.tcs.2018.06.036}}.

\bibitem{Nazeri2023}
Mahdi Nazeri, Ali Mollahosseini, and Iman Izadi.
\newblock A centrality based genetic algorithm for the graph burning problem.
\newblock {\em Applied Soft Computing}, 144:110493, 2023.
\newblock \href {https://doi.org/10.1016/j.asoc.2023.110493}
  {\path{doi:10.1016/j.asoc.2023.110493}}.

\bibitem{Pereira2021thesis}
F.~C. Pereira.
\newblock {A computational study of the Perfect Awareness Problem}.
\newblock Master's thesis, University of Campinas, Brazil, 2021.
\newblock URL: \url{https://hdl.handle.net/20.500.12733/1641217}.

\bibitem{Pereira2023}
F.~C. Pereira and P.~J. {de Rezende}.
\newblock {The Least Cost Directed Perfect Awareness Problem: complexity,
  algorithms and computations}.
\newblock {\em Online Social Networks and Media}, 37--38, 2023.
\newblock \href {https://doi.org/10.1016/j.osnem.2023.100255}
  {\path{doi:10.1016/j.osnem.2023.100255}}.

\bibitem{Pereira2021lagos}
F.~C. Pereira, P.~J. {de Rezende}, and C.~C. {de Souza}.
\newblock Effective heuristics for the perfect awareness problem.
\newblock {\em Procedia Computer Science}, 195:489--498, 2021.
\newblock \href {https://doi.org/10.1016/j.procs.2021.11.059}
  {\path{doi:10.1016/j.procs.2021.11.059}}.

\bibitem{complementary-data}
F.~C. Pereira, P.~J. {de Rezende}, and Tallys Yunes.
\newblock {A Row Generation Algorithm for Finding Optimal Burning Sequences of
  Large Graphs - Complementary Data}.
\newblock Mendeley Data, V1, 2024.
\newblock \href {https://doi.org/10.17632/c95hp3m4mz}
  {\path{doi:10.17632/c95hp3m4mz}}.

\bibitem{network_reposity}
Ryan~A. Rossi and Nesreen~K. Ahmed.
\newblock The network data repository with interactive graph analytics and
  visualization.
\newblock In {\em AAAI}, 2015.
\newblock URL: \url{https://networkrepository.com}.

\bibitem{Shakarian2013}
Paulo Shakarian, Sean Eyre, and Damon Paulo.
\newblock A scalable heuristic for viral marketing under the tipping model.
\newblock {\em Social Network Analysis and Mining}, 3(4):1225--1248, 2013.
\newblock \href {https://doi.org/10.1007/s13278-013-0135-7}
  {\path{doi:10.1007/s13278-013-0135-7}}.

\bibitem{Sim2018}
Kai~An Sim, Ta~Sheng Tan, and Kok~Bin Wong.
\newblock On the burning number of generalized petersen graphs.
\newblock {\em Bulletin of the Malaysian Mathematical Sciences Society},
  41(3):1657--1670, 2018.
\newblock \href {https://doi.org/10.1007/s40840-017-0585-6}
  {\path{doi:10.1007/s40840-017-0585-6}}.

\bibitem{Tan2023}
Ta~Sheng Tan and Wen~Chean Teh.
\newblock Burnability of double spiders and path forests.
\newblock {\em Applied Mathematics and Computation}, 438:127574, 2023.
\newblock \href {https://doi.org/https://doi.org/10.1016/j.amc.2022.127574}
  {\path{doi:https://doi.org/10.1016/j.amc.2022.127574}}.

\bibitem{Simon2019b}
Marek Šimon, Ladislav Huraj, Iveta Dirgova~Luptakova, and Jiri Pospichal.
\newblock How to burn a network or spread alarm.
\newblock {\em MENDEL}, 25(2):11--18, 2019.
\newblock \href {https://doi.org/10.13164/mendel.2019.2.011}
  {\path{doi:10.13164/mendel.2019.2.011}}.

\bibitem{Simon2019a}
Marek Šimon, Ladislav Huraj, Iveta Dirgová~Luptáková, and Jiří
  Pospíchal.
\newblock Heuristics for spreading alarm throughout a network.
\newblock {\em Applied Sciences}, 9(16), 2019.
\newblock \href {https://doi.org/10.3390/app9163269}
  {\path{doi:10.3390/app9163269}}.

\end{thebibliography}

\end{document}